\documentclass[aps,prl,reprint,groupedaddress]{revtex4-1}

\usepackage{amsmath,graphicx,color}
\usepackage{bm}

\newcommand{{\todo}}[1]{{\color{red} \bf #1}}

\begin{document}

\title{Topological superconductivity induced by a triple-Q magnetic structure}

\author{Jasmin Bedow$^{1,2}$, Eric Mascot$^{1}$, Thore Posske$^{3}$,  G\"{o}tz S. Uhrig$^{2}$, Roland Wiesendanger$^{4}$, Stephan Rachel$^{5}$, and Dirk K. Morr$^{1}$}
\affiliation{$^{1}$ University of Illinois at Chicago, Chicago, IL 60607, USA \\
$^{2}$ Lehrstuhl f\"ur Theoretische Physik I, TU Dortmund University, 44221 Dortmund, Germany \\
$^{3}$ I. Institut f\"{u}r Theoretische Physik, Universit\"{a}t Hamburg, 20355 Hamburg, Germany\\
$^{4}$ Department of Physics, Universit\"{a}t Hamburg, 20355 Hamburg, Germany\\
$^{5}$ School of Physics,  University of Melbourne, Parkville, VIC 3010, Australia
}

\date{\today}

\begin{abstract}
We demonstrate that the recently discovered triple-Q (3Q) magnetic structure, when embedded in a magnet-superconductor hybrid (MSH) system, gives rise to the emergence of topological superconductivity.  We investigate the structure of chiral Majorana edge modes at domain walls, and show that they can be distinguished from trivial in-gap modes through the spatial distribution of the induced supercurrents. Finally, we show that topological superconductivity in 3Q MSH systems is a robust phenomenon that does not depend on the relative alignment of the magnetic and superconducting layers, or on the presence of electronic degrees of freedom in the magnetic layer.

\end{abstract}

\maketitle

{\bf Introduction} The non-Abelian braiding statistics of Majorana zero modes (MZMs) represent a new paradigm for the realization of topological quantum computing and topology-based devices \cite{Nayak2008}. While these modes have been observed in one- \cite{Mourik2012,Das2012,Ruby2015,Pawlak2016,Kim2018} and two-dimensional (2D) \cite{Menard2017,Palacio-Morales2019,Menard2019,Machida2019} topological superconductors, their engineering at the atomic level and unambiguous experimental identification have remained an outstanding problem.  Magnet-superconductor hybrid (MSH) systems have emerged as a promising class of materials to overcome this problem as they allow (a) for the design of topological superconductivity using atomic manipulation \cite{Kim2018} and interface engineering techniques \cite{Palacio-Morales2019,Rachel2017}, and (b) to investigate Majorana modes using scanning tunneling spectroscopy (STS). MSH systems with non-collinear magnetic structures, such as magnetic skyrmions \cite{Heinze2011,Romming2015}, have become of great interest as they might not only exhibit topological superconductivity \cite{Nakosai2013,Yang2016,Rex2019,Garnier2019,Mascot2020} even in the absence of a Rashba spin-orbit interaction, but also provide the experimental ability to easily tune between different topological phases \cite{Mascot2020}. The recent experimental breakthrough in creating such a MSH system by depositing a non-collinear triple-Q (3Q) magnetic structure \cite{Kurz2001} on the surface of an $s$-wave superconductor \cite{Spethmann2020} in Mn/Re(0001), has raised the intriguing question whether this system will exhibit topological superconductivity below the superconducting critical temperature. Moreover, the observation of domain walls in Mn/Re(0001) provides an unprecedented opportunity to investigate what type of domain wall -- electronic, magnetic or structural -- is best suited to engineer Majorana modes, and how to distinguish them from trivial in-gap states, an important outstanding question for the unambiguous identification of topological states.

In this Letter, we address these open questions and demonstrate that MSH systems containing a 3Q magnetic layer do not only realize topological superconductivity, but also exhibit a rich topological phase diagram. Underlying the emergence of topological superconductivity is a uniform Rashba spin-orbit interaction that is induced by the 3Q magnetic structure. We show that ribbons of the 3Q magnetic structure exhibit Majorana edge modes, that despite the complex 3Q structure, still exhibit a well-defined chirality. Moreover, we demonstrate that domain walls at which the magnetic 3Q structure is inverted (spin-domain wall) give rise to Majorana modes that traverse the superconducting gap. In contrast, domain walls at which the superconducting order parameter undergoes a $\pi$-phase shift ($\pi$-phase domain wall) induce only trivial in-gap modes. However, while the local density of states (LDOS) at these two types of domain walls shows only some quantitative differences, we demonstrate that the induced supercurrents at the domain walls, which can be imaged using a scanning superconducting quantum interference device (SQUID) \cite{Spanton2014}, provide a qualitative signature that allows one to differentiate between trivial and topological in-gap states. Finally, we show that topological superconductivity in 3Q MSH systems is a robust phenomenon that does not depend on the relative alignment of the magnetic and superconducting layers, or on the presence of electronic degrees of freedom in the magnetic layer. Our results provide new guidance for the engineering and unambiguous experimental identification of chiral Majorana edge modes in general and for the search of Majorana modes in the recently observed triple-Q structure in Mn/Re(0001) in particular.

\begin{figure}
  \centering
  \includegraphics[width=8.5cm]{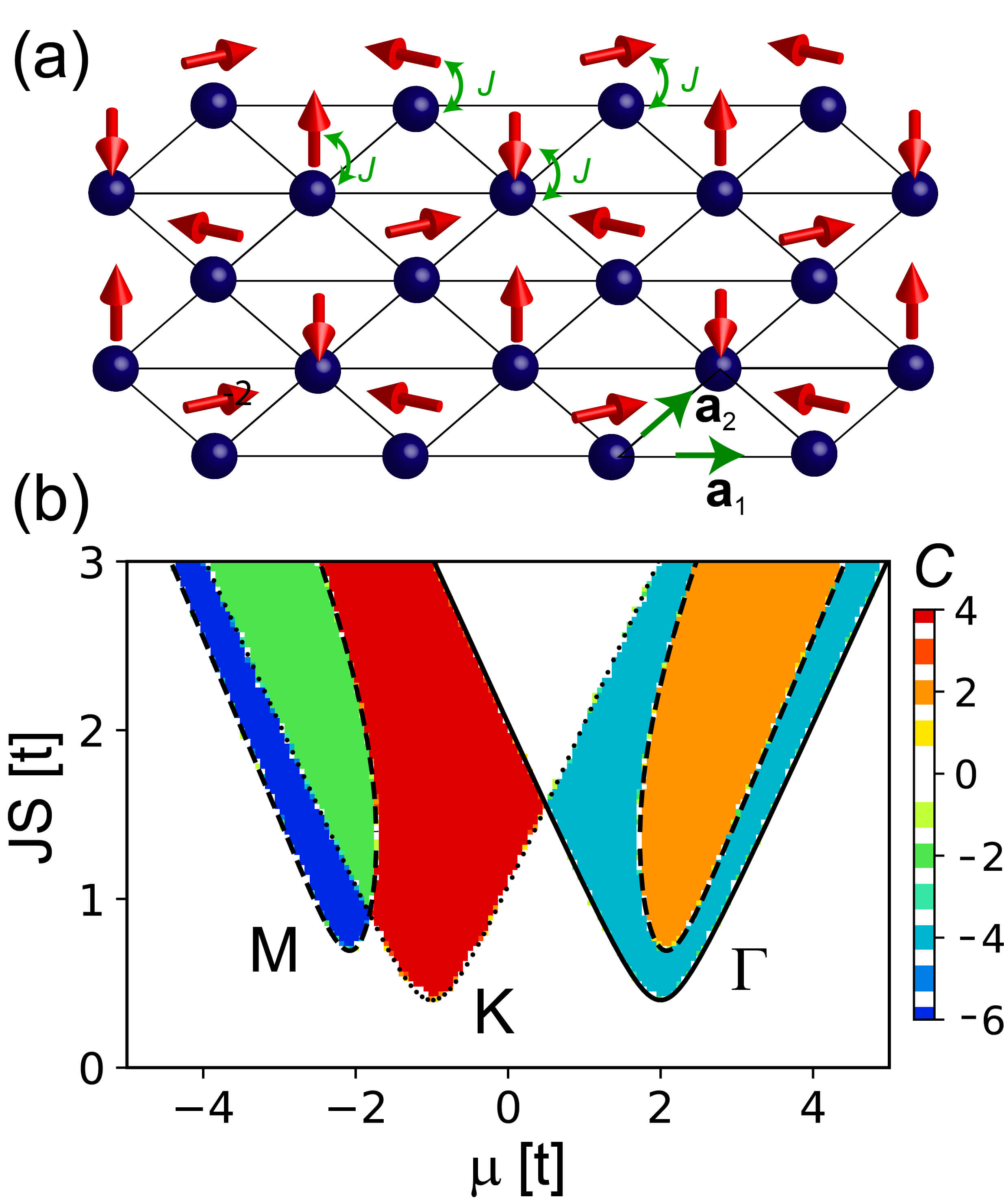}
  \caption{(a) Schematic picture of the 3Q magnetic layer (red arrows) placed on the surface of an $s$-wave superconductor (blue spheres). (b) Topological phase diagram representing the Chern number in the $(\mu,JS)$-plane for a 3Q MSH system with $\Delta = 0.4 t$. The solid, dashed and dotted lines denote gap closings at the $\Gamma$, $M$, and $(K,K^\prime)$ points, respectively. }
  \label{fig:Fig1}
\end{figure}

{\bf Theoretical Model}
To study the topological phase diagram of an MSH system containing a 3Q magnetic layer, we begin by considering a model, in which the magnetic moments are located above the sites of the underlying $s$-wave superconductor, and couple to the surface electrons via an exchange field only [see Fig.\ref{fig:Fig1}(a)] (a more complex model motivated by recent experiments \cite{Kim2018,Palacio-Morales2019}, will be discussed below). Such a system is described by the Hamiltonian $\mathcal{H}_1 = \mathcal{H}_0 + \mathcal{H}_m$ with
\begin{align}
    \mathcal{H}_0 &= - \sum_{\langle {\bf r}, {\bf r^\prime}\rangle ,\sigma} t_{{\bf r}{\bf r}^\prime } c^\dagger_{{\bf r}\sigma} c_{\bf{r^\prime}\sigma} -\mu \sum_{{\bf r},\sigma} c^\dagger_{{\bf r}\sigma} c_{{\bf r}\sigma} \nonumber \\
    & + \sum_{{\bf r}} \left(  \Delta_0 c^\dagger_{{\bf r}\uparrow} c^\dagger_{{\bf r}\downarrow}  + H.c. \right) \; , \nonumber\\
    \mathcal{H}_m &= \sum_{{\bf r},\alpha,\beta} c^\dagger_{{\bf r}\alpha} ({\bf S}_{{\bf r}} \cdot {\bm \sigma})_{\alpha\beta} c_{{\bf r}\beta} \; ,
    \label{eq:H}
\end{align}
where $- t_{{\bf r}{\bf r}^\prime }$ is the hopping amplitude between nearest neighbor sites on a triangular lattice, $\mu$ is the chemical potential, and $\Delta_0$ is the superconducting $s$-wave order parameter. $J$ is the magnetic exchange coupling, and $c^\dagger_{{\bf r}\sigma}$ creates an electron with spin $\sigma$ at site ${\bf r}$. ${\bf S}_{{\bf r}}$ represents the magnetic moment's spin $S$ at site ${\bf r}$, encoding the spatial form of the 3Q magnetic  structure. It possesses a $2 \times 2$-unit cell with the four spins pointing from the center of a tetrahedron to its corners [see Fig.~\ref{fig:Fig1}(a) and supplemental information (SI) section 1].  As the hard superconducting gap suppresses Kondo screening, we consider $\mathbf{S}_{\mathbf{r}}$ to represent classical spins. Due to the particle-hole symmetry of the superconducting state, and the broken time-reversal symmetry arising from the presence of magnetic moments, the topological superconductor belongs to the topological class D \cite{Kitaev2009,Ryu2010}.

To characterize the topological superconducting phases of the system, we compute the topological invariant -- the Chern number $C$ -- given by
\begin{align}\label{eq:C}
 C & =  \frac{1}{2\pi i} \int_{\text{BZ}} d^2k \mathrm{Tr} ( P_{\bf{k}} [ \partial_{k_x} P_{\bf{k}}, \partial_{k_y} P_{\bf{k}} ] )  \ ;  \nonumber \\
 P_{\bf{k}} & = \sum_{E_n(\bf{k}) < 0} |\Psi_n({\bf{k}}) \rangle \langle \Psi_n({\bf{k}})| \ ,
\end{align}
where $E_n({\bf{k}})$ and $|\Psi_n({\bf{k}}) \rangle$ are the eigenenergies and the eigenvectors of the Hamiltonian in Eq.(\ref{eq:H}), with $n$ being a band index, and the trace is taken over Nambu and spin-space.

{\bf Topological Phase Diagram}
In addition to a hard superconducting gap and a magnetic order, the emergence of topological superconductivity in MSH systems usually requires the presence of a Rashba spin-orbit (RSO) interaction \cite{Li2016,Rachel2017}, which is absent in the Hamiltonian of Eq.(\ref{eq:H}). However, an effective RSO interaction is induced by the non-collinear 3Q magnetic structure, as can be seen by applying a local unitary transformation to Eq.(\ref{eq:H}) that rotates the local spin into the $z$-axis. This transformation results in a ferromagnetic spin order, perpendicular to the layer, and a RSO interaction with spatially uniform magnitude $|\alpha| = \sqrt{2 \:/\: 3} \, t$ (see SI section 1), thus satisfying all requirements for the emergence of topological superconductivity.

The topological phase diagram in the ($\mu,JS$)-plane, resulting from the Hamiltonian in Eq.(\ref{eq:H}), exhibits a rich structure [see Fig.~\ref{fig:Fig1}(b)]. The topological phases possess only even-numbered Chern numbers, which is a direct consequence of the symmetry of the 3Q magnetic structure that leads to doubly degenerate electronic bands (see SI section 2). The phase boundaries are determined by the closing of the gap at the time-reversal invariant $\Gamma, M, (K,K^\prime)$ points in the Brillouin zone (see solid, dashed and dotted lines respectively, in Fig.~\ref{fig:Fig1}(b)], and are determined by the conditions
\begin{align}
 \mu & =  \pm \sqrt{J^2 - \Delta^2} +\mu_0
 \end{align}
 with $\mu_0 = 2t$ for the $\Gamma$ point, and $\mu_0 = -t$ at the $(K,K^\prime)$ points, and
\begin{align}
\mu & = \pm \sqrt{J^2-\Delta^2+4t^2 \pm \sqrt{\frac{16}{3}(J^2 - \Delta^2)t^2} }
\end{align}
at the $M$ point. While the gap closing at the $\Gamma$ and $(K,K^\prime)$ points yields phase transition lines that are symmetric around $\mu=0.5$t, those at the $M$ point are symmetric around $\mu=0$. As a result, the topological phase with $C=2$ (for $\mu>0$) is fully surrounded by one with $C=-4$, while their counterparts (for $\mu<0$) only partly overlap, giving rise not only to $C=-2$ and $C=4$ phases, but also to a $C=-6$ phase. We note that the gap closing at the $(K,K^\prime)$ and $M$ points occur via a Dirac cone, which together with a multiplicity of the $(K,K^\prime)$ and $M$ points of $m=4,6$, respectively, leads to a change in the Chern number by $\Delta C = 4,6$. In contrast, the gap closing at the $\Gamma$ point leads to a quadratic dispersion, $E_k \sim \pm k^2$, which together with its multiplicity of $m=2$ implies a change in the Chern number by $\Delta C = 4$ at the phase transition. Note that the phase diagram is invariant under uniform rotations of the spin structure (up to an overall change in the sign of $C$).

\begin{figure}[t]
  \centering
  \includegraphics[width=8.5cm]{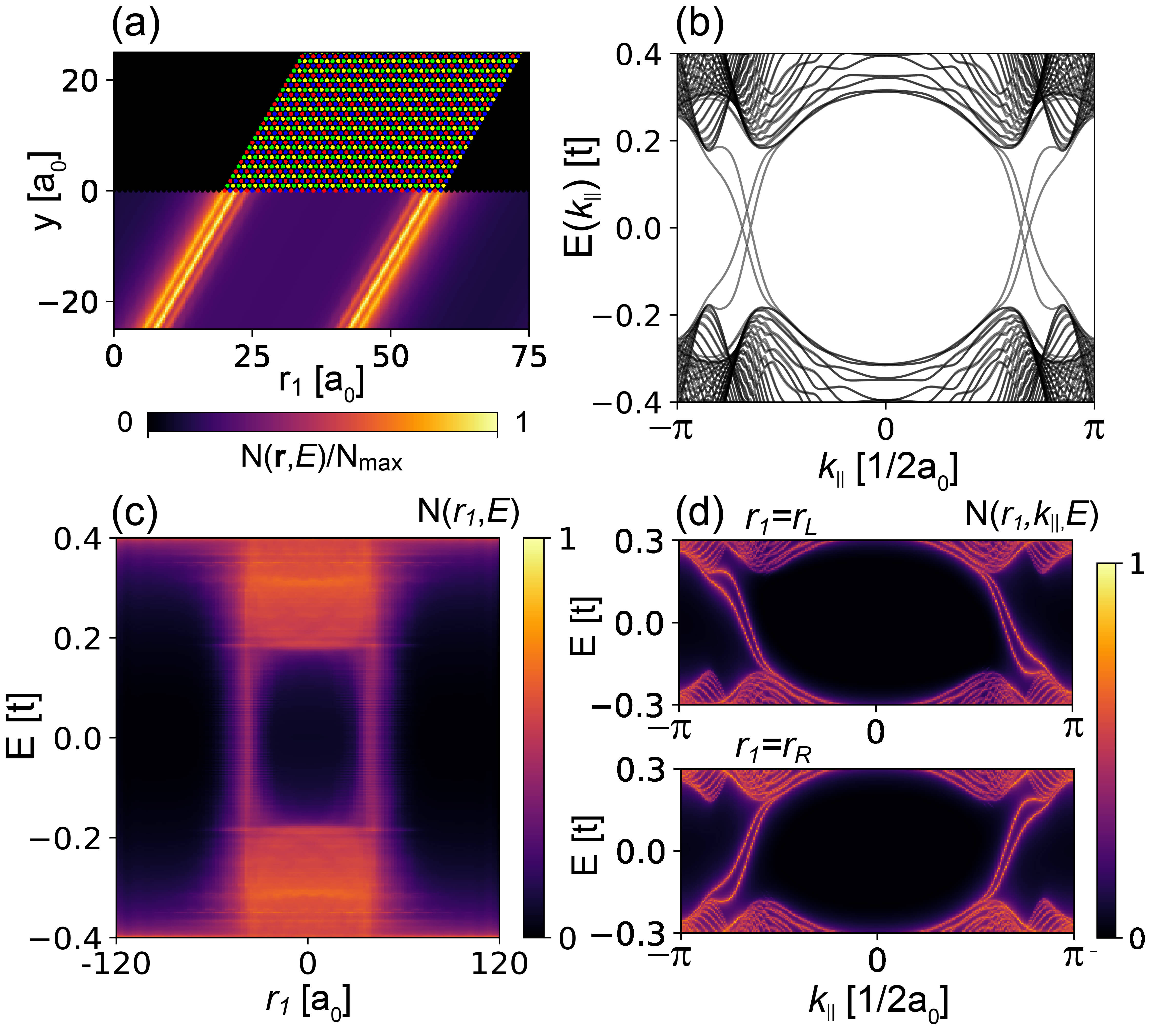}
  \caption{(a) Spatial plot for a 3Q ribbon periodic along ${\bf a}_2$ with width $L=40a_0$ along ${\bf a}_1$, located on the surface of the $s$-wave superconductor (upper panel), and the zero-energy LDOS (lower panel) showing Majorana edge modes. (b) Electronic dispersion as a function of momentum $k_\parallel$ along the ribbon. (c) Line-cut (along ${\bf a}_1$) through the ribbon of the energy resolved LDOS. (d) LDOS at the left (upper panel) and right edges (lower panel) of the ribbon, momentum resolved along the ribbon, i.e., $N(r_1, k_\parallel, E)$. The ribbon is in the $C=4$ phase with parameters $(\mu,\Delta,JS)=(-1,0.4,1)t$.}
  \label{fig:Fig2}
\end{figure}

{\bf Ribbon geometry}
In order to investigate the emergence of Majorana edge modes in 3Q MSH systems, we next consider a ribbon of the 3Q magnetic layer placed on the surface of an $s$-wave superconductor, as shown in the upper panel of Fig.~\ref{fig:Fig2}(a). In a topological phase with Chern number $C$, adjacent to a trivial phase, the bulk-boundary correspondence requires that each edge possesses $|C|$ chiral Majorana edge modes \cite{Bernevig2013}. These modes traverse the superconducting gap and disperse linearly near the Fermi energy as a function of the momentum along the ribbon edge, as shown in  Fig.~\ref{fig:Fig2}(b) for the $C=4$ phase.
A plot of the zero-energy local-density of states (LDOS), [see lower panel in Fig.~\ref{fig:Fig2}(a)] reveals that the Majorana modes are as expected localized at the edges of the 3Q ribbon. The Majorana modes remain localized at the ribbon's edges with increasing energy, as shown in Fig.~\ref{fig:Fig2}(c), where we present a linecut of the energy-resolved LDOS, $N({\bf r},E)$, along ${\bf a}_1$ through the ribbon. However, with increasing energy, the modes extend further away from the edges which is a direct result of the mode's localization length which increases with increasing energy and diverges at the band edges. Fig.~\ref{fig:Fig2}(c) also reveals that the superconducting gap inside the ribbon is suppressed by the presence of the 3Q magnetic structure \cite{Rachel2017}.  Moreover, considering the LDOS momentum resolved along the ribbon and summed over both spin states, i.e., $N(r_1, k_\parallel, E)$ [see Fig.~\ref{fig:Fig2}(d)], we find that all four Majorana modes at the right edge ($r_1 = r_{R}$) are right movers ($\partial E/\partial k_\parallel >0$) while those at the left edge ($r_1 = r_{L}$) are left movers ($\partial E/\partial k_\parallel <0$). This implies, that despite the complex magnetic structure of the 3Q layer, the Majorana edge modes still possess a well-defined chirality.

\begin{figure*}[t]
  \centering
  \includegraphics[width=17.0cm]{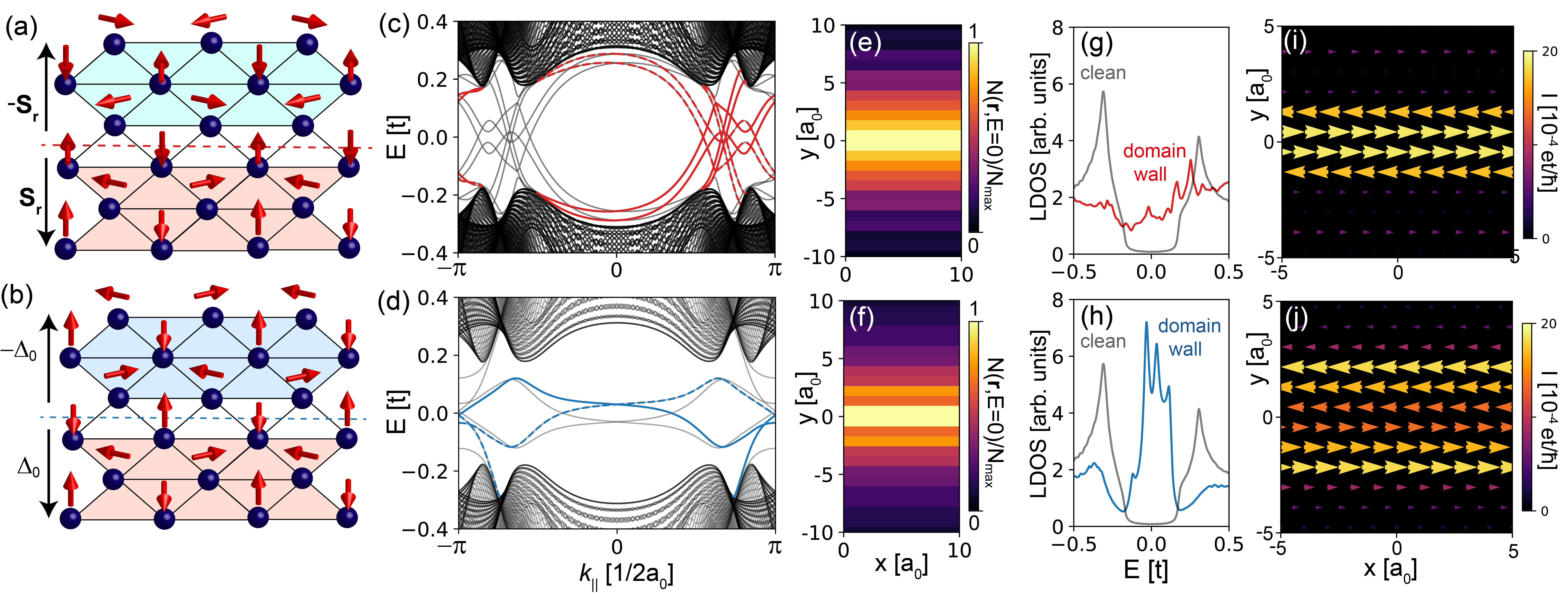}
  \caption{Schematic picture of (a) a spin domain wall where ${\bf S}_{\bf r} \rightarrow -{\bf S}_{\bf r}$ (case I), and (b) $\pi$-phase domain wall with superconducting order parameter $\Delta_0 \rightarrow - \Delta_0$ (case II). Electronic structure as a function of momentum $k_\parallel$ along the domain wall for (c) case I (Majorana modes are shown as red lines), and (d) case II (trivial in-gap states shown in blue). Zero energy LDOS for (e) case I, and (f) case II. LDOS at the domain wall for (g) case I, and (h) case II. Spatial structure of the supercurrent near the domain wall for (i) case I, and (j) case II. Parameters are $(\mu, \Delta, JS) = (-1.0, 0.4, 1.0)t$ yielding a $C=4$ phase.}
  \label{fig:Fig3}
\end{figure*}
{\bf Domain walls}
The recent observation of a dispersing Majorana mode at a domain wall \cite{Wang2020} in the putative topological superconductor FeSe$_{0.45}$Te$_{0.55}$ \cite{Wang2018,Machida2019} raises the question of what types of domain walls induce chiral Majorana modes. To investigate this question, we consider two different types of domain walls: a spin domain wall at which the spin-structure of the 3Q layer is inverted, i.e., ${\bf S}_{\bf r} \rightarrow -{\bf S}_{\bf r}$ [see Fig.~\ref{fig:Fig3}(a), case I], and a $\pi$-phase domain wall at which the superconducting order parameter experiences a $\pi$-phase shift, i.e., $\Delta_0 \rightarrow - \Delta_0$ [see Fig.~\ref{fig:Fig3}(b), case II]. For a spin domain wall, the domains possess Chern numbers $C$ and $-C$, leading to the emergence of $\Delta C = 2|C|$ Majorana modes at the domain wall (here, $C=4$). The modes traverse the superconducting gap, as shown in Fig.~\ref{fig:Fig3}(c) (see red solid and dashed lines), where we present the electronic dispersion as a function of momentum $k_\parallel$ along the domain wall. As we employ periodic boundary conditions, our system possesses two domain walls, leading to a total of $4|C|$ Majorana modes in the electronic dispersion. In contrast, for a $\pi$-phase domain wall, the domains possess the same Chern number, and hence the bulk-boundary correspondence does not require the emergence of Majorana modes. Indeed, while the $\pi$-phase domain wall induces in-gap modes [see Fig.~\ref{fig:Fig3}(d)], they do not connect the upper and lower bands, and are thus not topological. However, for both domain walls, the in-gap states are localized near the domain boundary, as follows from the plot of the zero-energy LDOS, $N({\bf r},E=0)$, shown in Figs.~\ref{fig:Fig3}(e) and \ref{fig:Fig3}(f) for cases I and II, respectively. In addition, a plot of the LDOS at the domain wall reveals substantial spectral weight inside the gap for both cases, with a nearly flat LDOS for case I [see Fig.~\ref{fig:Fig3}(g)], and a strong peak near zero energy for case II [see Fig.~\ref{fig:Fig3}(h)]. Therefore, it is difficult to experimentally distinguish these two cases, and the ensuing topological or trivial character of the induced in-gap states, based solely on the LDOS.

However, a significant and qualitative difference between spin and $\pi$-phase domain walls emerges when considering the supercurrents induced by the domain walls. For a spin domain wall, the opposite Chern numbers imply that the chirality of the supercurrents associated with the two domains is reversed. As a result, the supercurrents posses an even symmetry across the domain wall, leading to a non-vanishing net supercurrent flowing along the domain wall [see Fig.~\ref{fig:Fig3}(i)].
In contrast, for a $\pi$-phase domain wall, the domains possess the same Chern number, and their associated supercurrents have the same chirality. As a result, the supercurrents possess an odd symmetry across the domain wall, and the net supercurrent along the domain wall vanishes. We thus conclude that by imaging the net supercurrent flowing along the domain wall, for example, using a SQUID \cite{Spanton2014}, it is possible to distinguish the existence of Majorana modes, as in case I, from the presence of trivial in-gap states, as in case II.

\begin{figure}
  \centering
  \includegraphics[width=8.5cm]{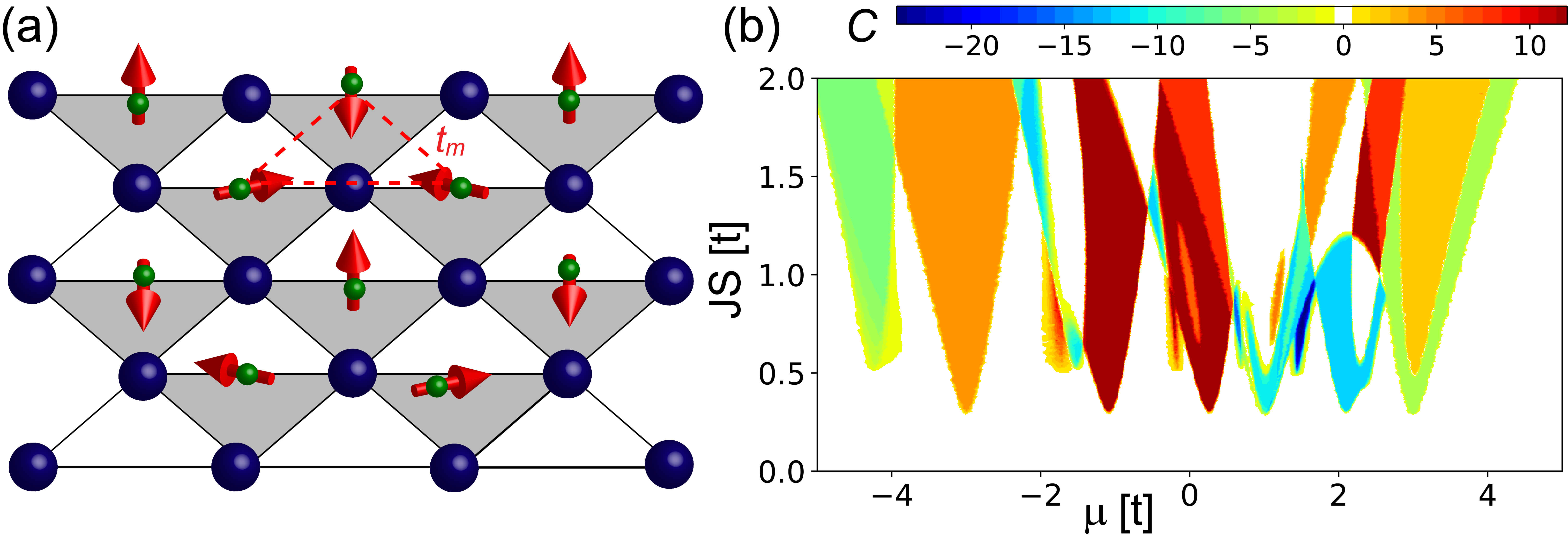}
  \caption{(a) 3Q magnetic layer with electronic degrees of freedom and adatoms located above the surface's hollow sites (shaded triangles). (b) Topological phase diagram in the $(\mu,JS)$-plane, with $t_{hyb}=t_m=t$, $\mu_m=\mu$ and $\Delta=0.3t$.}
  \label{fig:Fig4}
\end{figure}

{\bf Experimentally motivated model} While the above model has demonstrated the general existence of topological superconductivity in 3Q MSH systems, the question naturally arises of how this result depends on  material specifics. In particular, recent realizations of MSH systems have shown \cite{Kim2018,Palacio-Morales2019} that the magnetic adatoms reside in the hollow sites of the underlying superconducting substrate lattice [shaded triangles in Fig. 4(a)], in contrast to the assumption of on-top adsorption sites made so far [Fig.\ref{fig:Fig1}(a)]. Moreover, the magnetic layer itself possesses an electronic structure, which interacts with the magnetic moments through an exchange field and is coupled to the superconducting surface layer via electronic hopping. Such a system is described by the Hamiltonian $\mathcal{H} = \mathcal{H}_0 + \mathcal{H}^\prime_m+ \mathcal{H}_{hyb}$, where
\begin{align}
    \mathcal{H}^\prime_m &= -t_m \sum_{{\bf r}, {\bf r^\prime} \in M ,\sigma} d^\dagger_{{\bf r}\sigma} d_{\bf{r^\prime}\sigma} -\mu_m \sum_{{\bf r} \in M ,\sigma} d^\dagger_{{\bf r}\sigma} d_{{\bf r}\sigma} \nonumber \\
    & + \sum_{{\bf r} \in M,\alpha,\beta} d^\dagger_{{\bf r}\alpha} ({\bf S}_{{\bf r}} \cdot {\bm \sigma})_{\alpha\beta} d_{{\bf r}\beta} \ ,
    \label{eq:H2}
\end{align}
$\mathcal{H}_{hyb} = - t_{hyb} \sum_{ {\bf r}, {\bf r'}, \sigma} \left( c^\dagger_{{\bf r}\sigma} d_{{\bf r}^\prime\sigma} + H.c. \right)$ and $\mathcal{H}_0$ given in Eq.(\ref{eq:H}). Here, $d^\dagger_{{\bf r}\sigma}$ creates an electron with spin $\sigma$ at site ${\bf r}$ in the magnetic layer, the sum in $\mathcal{H}^\prime_m$ runs over all sites of the triangular magnetic lattice $M$,
and $\mathcal{H}_{hyb}$ describes the electronic hopping between a site in the magnetic layer and its nearest neighbor sites on the superconducting surface. The resulting topological phase diagram in the $(\mu,JS)$-plane [see Fig.~\ref{fig:Fig4}(b)] reveals not only a significantly larger number of topological phases than shown in Fig.~\ref{fig:Fig1}(b), but also larger magnitudes of the Chern number. These results are robust against changes in the band parameters (see SI section 3). As before, the electronic bands of the system are doubly degenerate, resulting in topological phases with even Chern number only.

{\bf Conclusions} We have shown that a non-collinear 3Q magnetic layer in a MSH system induce a spatially uniform Rashba spin-orbit interaction that gives rise to a rich topological phase diagram. We demonstrated that a spin domain leads to topological Majorana modes, while a $\pi$-phase domain wall induces trivial in-gap states only.  Both types of domain walls possess a qualitatively similar LDOS,  making it difficult to distinguish between them experimentally solely based on STS measurements. In contrast, the spatial distribution of the induced supercurrents differs qualitatively, allowing one to distinguish topological domain walls giving rise to Majorana modes from trivial ones. Finally, we show that the emergence of topological phases in the 3Q MSH system is a robust phenomenon that does not depend on the relative alignment of the magnetic and superconducting layers, or the presence of electronic degrees of freedom in the magnetic layer. Our results open new possibilities for the creation of topological superconductivity and the manipulation of the concomitant edge modes in MSH systems containing complex magnetic structures. The long-term vision is to control and to modify electronic circuits of edge modes by manipulating domain walls as proposed for magnonic edge modes \cite{Wang2018a}.

\section{Acknowledgements}

The authors would like to thank H.~Kim, A.~Kubetzka, and K.~von Bergmann for stimulating discussions. This work was supported by the U. S. Department of Energy, Office of Science, Basic Energy Sciences, under Award No. DE-FG02-05ER46225 (E.M.,D.K.M.), ERC Advanced Grant ADMIRE (project No. 786020) (R.W.), the {\it Studienstiftung des deutschen Volkes} (J.B.), and through ARC DP200101118 (S.R.).\\

\end{document}

\title{{\Large Topological superconductivity induced by a triple-Q magnetic structure\\[0.5cm]}
{\large Supplemental Information}}

\author{Jasmin Bedow$^{1,2}$, Eric Mascot$^{1}$, Thore Posske$^{3}$,  G\"{o}tz S. Uhrig$^{2}$, Roland Wiesendanger$^{4}$, Stephan Rachel$^{5}$, and Dirk K. Morr$^{1}$}
\affiliation{$^{1}$ University of Illinois at Chicago, Chicago, IL 60607, USA \\
$^{2}$ Lehrstuhl f\"ur Theoretische Physik I, TU Dortmund University, 44221 Dortmund, Germany \\
$^{3}$ I. Institut f\"{u}r Theoretische Physik, Universit\"{a}t Hamburg, 20355 Hamburg, Germany\\
$^{4}$ Department of Physics, Universit\"{a}t Hamburg, 20355 Hamburg, Germany\\
$^{5}$ School of Physics,  University of Melbourne, Parkville, VIC 3010, Australia
}

\maketitle

\begin{center}
{\bf Section 1: Rashba spin-orbit interaction induced by the 3Q magnetic structure}
\end{center}

\begin{figure}[h]
  \centering
  \includegraphics[width=8.5cm]{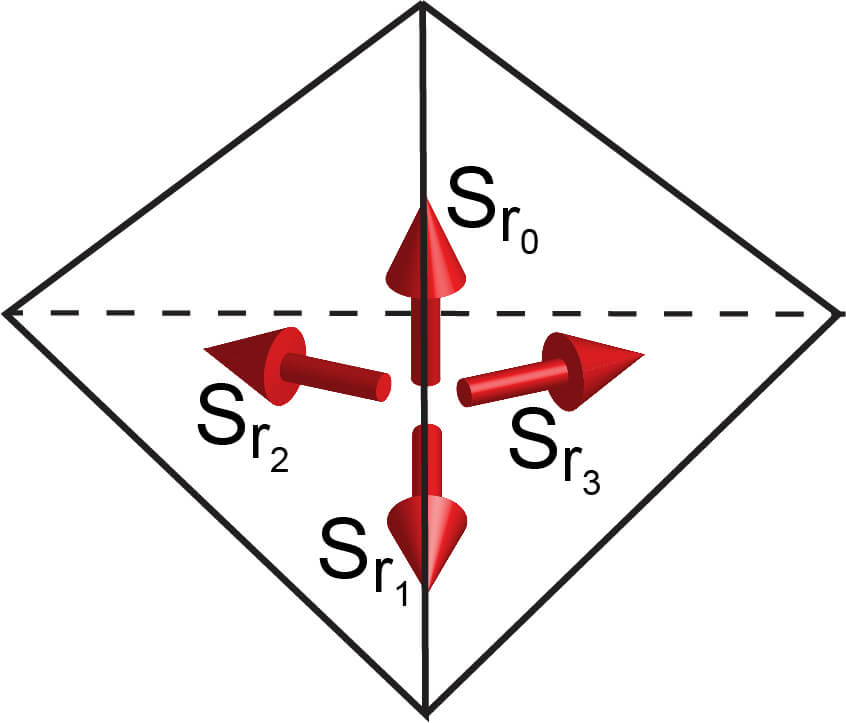}
  \caption{The four spins $\mathbf{S}_{\mathbf{r}_i}$ of the 3Q magnetic structure's $2 \times 2$-unit cell pointing from the center of a tetrahedron to its corners}
  \label{fig:SI_Fig1}
\end{figure}

The unit cell of the 3Q magnetic structure is a $2 \times 2$-unit cell with the four spins $\mathbf{S}_{\mathbf{r}_i}$ pointing from the center of a tetrahedron to its corners, as shown in Fig.~\ref{fig:SI_Fig1} and described by
\begin{equation}
  \mathbf{S}_{\mathbf{r}_i} =
  \begin{cases}
    S (0,0,1) , & \mathbf{r}_0 = \mathbf{0} \\
    S (0, -\sqrt{8}/3, -1/3), & \mathbf{r}_1 = \mathbf{a}_1 \\
    S (-\sqrt{6}/3, \sqrt{2}/3, -1/3), & \mathbf{r}_2 = \mathbf{a}_2 \\
    S (\sqrt{6}/3, \sqrt{2}/3, -1/3), & \mathbf{r}_3 = \mathbf{a}_1 + \mathbf{a}_2   \; ,
  \end{cases}
\end{equation}
with $\mathbf{r}_i$ denoting the atom's position in the unit cell with lattice vectors ${\bf a}_1=(1,0)a_0$ and ${\bf a}_2=(1/2,\sqrt{3}/2)a_0$ and lattice constant $a_0$.

To demonstrate that the 3Q magnetic structure induces an effective Rashba spin-orbit interaction, we perform a local unitary transformation that rotates the local spin $\vec{S}_{\vec{r}}$ into the ${\hat z}$-axis, perpendicular to the plane of the MSH system. The corresponding unitary transformation is defined via
\begin{equation}
	\begin{pmatrix}
		c_{\vec{r} \uparrow} \\
		c_{\vec{r} \downarrow}
	\end{pmatrix}
	= {\hat U}_{\vec{r}}
	\begin{pmatrix}
		d_{\vec{r} \uparrow} \\
		d_{\vec{r} \downarrow} \; ,
	\end{pmatrix}
\end{equation}
where the unitary matrix ${\hat U}_{\vec{r}} $ is given by
\begin{align}
  U_{\bf r} = e^{i \frac{\theta_{\bf r}}{2} \sigma \cdot \hat{n}_{\bf r}}
\end{align}
with
\begin{align}
   \cos(\theta_{\bf r}) &= \frac{S_{\bf r}}{|S_{\bf r}|} \cdot \hat{z} \\
   \intertext{and}
   \hat{n}_{\bf r} &= \frac{S_{\bf r} \times \hat{z}}{|S_{\bf r} \times \hat{z}|} \; .
\end{align}

The Hamiltonian of Eq.(1) of the main text in this new basis then takes the form
\begin{align}
	H =& \sum_{\vec{r}, \vec{r}', \alpha, \beta} \left(
		-t_{\vec{r}\vec{r}'} {\hat U}_{\vec{r}}^\dagger {\hat U}_{\vec{r}^\prime}
	\right)_{\alpha\beta} d_{\vec{r} \alpha}^\dagger d_{\vec{r} \beta}
	+ \Delta \sum_{\vec{r}} \left(
		d_{\vec{r} \uparrow}^\dagger d_{\vec{r} \downarrow}^\dagger
		+ \mathrm{H.c.}
	\right) + \sum_{\vec{r}, \alpha} (JS \sigma_{\alpha\alpha}^z - \mu) d_{\vec{r} \alpha}^\dagger d_{\vec{r} \alpha} \; .
\end{align}
The effective hopping is given by
\begin{equation}
	-{\hat t}_{\vec{r}\vec{r}'}=-t_{\vec{r}\vec{r}'} {\hat U}_{\vec{r}}^\dagger {\hat U}_{\vec{r}^\prime}
	= \begin{pmatrix}
		-\tilde{t}_{\vec{r}\vec{r}'} &
		-\alpha_{\vec{r}\vec{r}'}^* \\
		\alpha_{\vec{r}\vec{r}'} &
		-\tilde{t}_{\vec{r}\vec{r}'}^*
	\end{pmatrix} \ ,
\end{equation}
with \(\alpha_{\vec{r}\vec{r}'}\) being the induced Rashba spin-orbit interaction. To derive the explicit form of the effective hopping $-{\hat t}_{\vec{r}\vec{r}'}$, we note that for the four different spins in the 3Q structure, the local gauge transformations are given through
\begin{subequations}
\begin{align}
  U_{{\bf r}_0} &= {\bf 1}_{2\times2} \; ;
  &
  U_{{\bf r}_1} &=
  \left(
  \begin{array}{cc}
   \frac{1}{\sqrt{3}} & -i \sqrt{\frac{2}{3}} \\
   -i \sqrt{\frac{2}{3}} & \frac{1}{\sqrt{3}} \\
  \end{array}
  \right) \; ;  \\
  U_{{\bf r}_2} &=
  \left(
  \begin{array}{cccc}
   \frac{1}{\sqrt{3}} & \frac{1}{\sqrt{2}} + i \frac{1}{\sqrt{6}} \\
   -\frac{1}{\sqrt{2}} + i \frac{1}{\sqrt{6}} & \frac{1}{\sqrt{3}} \\
  \end{array}
  \right)  \; ;
  &
  U_{{\bf r}_3} &=
  \left(
  \begin{array}{cccc}
   \frac{1}{\sqrt{3}} & -\frac{1}{\sqrt{2}} + i \frac{1}{\sqrt{6}} \\
   \frac{1}{\sqrt{2}} + i \frac{1}{\sqrt{6}} & \frac{1}{\sqrt{3}} \\
  \end{array}
  \right) \; .
\end{align}
\end{subequations}
There exist six unique bonds that connect the four spins in the unit cell [see Fig.1(a) of the main text]. For these, the resulting hopping matrices are given as
\begin{subequations}
\begin{align}
  \hat{t}_{01} &=
  \left(
  \begin{array}{cc}
   \frac{1}{\sqrt{3}} & -i \sqrt{\frac{2}{3}} \\
   -i \sqrt{\frac{2}{3}} & \frac{1}{\sqrt{3}} \\
  \end{array}
  \right) \; ;
  &
  \hat{t}_{02} &=
  \left(
  \begin{array}{cc}
   \frac{1}{\sqrt{3}} & \frac{1}{\sqrt{2}} + i \frac{1}{\sqrt{6}} \\
   -\frac{1}{\sqrt{2}} + i \frac{1}{\sqrt{6}} & \frac{1}{\sqrt{3}} \\
  \end{array}
  \right) \; ;
  &
  \hat{t}_{03} &=
  \left(
  \begin{array}{cc}
   \frac{1}{\sqrt{3}} & -\frac{1}{\sqrt{2}} + i \frac{1}{\sqrt{6}} \\
   \frac{1}{\sqrt{2}} + i \frac{1}{\sqrt{6}} & \frac{1}{\sqrt{3}} \\
  \end{array}
  \right) \; ;
  \\
  \hat{t}_{12} &=
  \left(
  \begin{array}{cc}
   -\frac{i}{\sqrt{3}} & \frac{1}{\sqrt{6}}+\frac{i}{\sqrt{2}} \\
   -\frac{1}{\sqrt{6}}+\frac{i}{\sqrt{2}} & \frac{i}{\sqrt{3}} \\
  \end{array}
  \right) \; ;
  &
  \hat{t}_{13} &=
  \left(
  \begin{array}{cc}
   \frac{i}{\sqrt{3}} & -\frac{1}{\sqrt{6}}+\frac{i}{\sqrt{2}} \\
   \frac{1}{\sqrt{6}}+\frac{i}{\sqrt{2}} & -\frac{i}{\sqrt{3}} \\
  \end{array}
  \right) \; ;
  &
  \hat{t}_{23} &=
  \left(
  \begin{array}{cc}
   -\frac{i}{\sqrt{3}} & -\sqrt{\frac{2}{3}} \\
   \sqrt{\frac{2}{3}} & \frac{i}{\sqrt{3}} \\
  \end{array}
  \right) \; ,
\end{align}
\end{subequations}
with $\hat{t}_{ij} = \hat{t}^\dagger_{ji} = U_{{\bf r}_i}^\dagger U_{{\bf r}_j}$. Here, $\hat{t}_{ij}$ denotes the hopping matrix between sites ${\bf r}_i$ and ${\bf r}_j$. The above results show that while the phase of \(\alpha_{\vec{r}\vec{r}'}\) varies spatially, its magnitude is constant, $|\alpha| = \sqrt{2/3} t$. At the same time, the magnitude of the nearest-neighbor hopping term is reduced to $|\tilde{t}_{\vec{r}\vec{r}'}| = \sqrt{1/3} t$.\\

\begin{center}
{\bf Section 2: Symmetries of the magnetic 3Q structure and doubly degenerate electronic band structure}
\end{center}

The real space structure of the 3Q magnetic layer is shown in Fig.~1(a) of the main text. The symmetry group of the 3Q magnetic structure is the tetrahedral group, which includes three $C_2$ symmetries and four $C_3$ symmetries. The doubly degenerate electronic band structure is a result of the $C_2$ symmetries, arising from simultaneous spin rotations and spatial translations. It suffices to analyze one of these in the following. Consider, for example, a translation of the spin structure by the lattice vector ${\bf a}_1$ [see Fig.~1(a) of the main text]. To transform the system into its original form, the spins ${\bf S}_{{\bf r}_0}$ and ${\bf S}_{{\bf r}_1}$ as well as ${\bf S}_{{\bf r}_2}$ and ${\bf S}_{{\bf r}_3}$
need to be rotated into one another (see Fig.~\ref{fig:SI_Fig1}). This can be achieved by rotating all four spins by an angle $\pi$ around the axis given by ${\bf S}_{{\bf r}_0} + {\bf S}_{{\bf r}_1}$. These two symmetry operations of translation represented within the unit cell and rotation in spin-space can be represented by matrices ${\hat T}$ and ${\hat R}$ given by
\begin{subequations}
\begin{align}
 {\hat  T}&=
  \begin{pmatrix}
    0 & 1 & 0 & 0 \\
    1 & 0 & 0 & 0 \\
    0 & 0 & 0 & 1 \\
    0 & 0 & 1 & 0 \\
  \end{pmatrix}
  \otimes {\bf 1}_{4\times4} \; ;
  \\
  {\hat R}&= {\bf 1}_{4\times4} \otimes \exp\left(i \frac{\pi}{2} \tau_0 \otimes \left({\bf n} \cdot \vec{\sigma}\right) \right) = {\bf 1}_{4\times4} \otimes i \tau_0 \otimes \left({\bf n} \cdot \vec{\sigma}\right) =
  {\bf 1}_{4\times4} \otimes
  \begin{pmatrix}
  \frac{i}{\sqrt{3}} & -\sqrt{\frac{2}{3}} & 0 & 0\\
      \sqrt{\frac{2}{3}} & -\frac{i}{\sqrt{3}} & 0 & 0\\
      0 & 0 & \frac{i}{\sqrt{3}} & -\sqrt{\frac{2}{3}}\\
      0 & 0 & \sqrt{\frac{2}{3}} & -\frac{i}{\sqrt{3}}
  \end{pmatrix} \; .
\end{align}
\end{subequations}
Here, both ${\hat T}$ and ${\hat R}$ are $16 \times 16$ matrices that acts on a spinor representing the four sites in a unit cell, defined by
\begin{equation}
  c_{\vec{k}} =
  \begin{pmatrix}
    c_{\vec{k}, \vec{r}_0} \\
    c_{\vec{k}, \vec{r}_1} \\
    c_{\vec{k}, \vec{r}_2} \\
    c_{\vec{k}, \vec{r}_3} \\
  \end{pmatrix} \; ,
\end{equation}
where $c_{\vec{k} \vec{r}_i}$ is given by
\begin{equation}
  c_{\vec{k}, \vec{r}_i} =
  \begin{pmatrix}
    c_{\vec{k}, \vec{r}_i, \uparrow} \\
    c_{\vec{k}, \vec{r}_i, \downarrow} \\
    c^\dagger_{\vec{k}, \vec{r}_i, \downarrow} \\
    -c^\dagger_{\vec{k}, \vec{r}_i, \uparrow} \\
  \end{pmatrix}
  \; .
\end{equation}
We note that a translation of the magnetic structure by ${\bf a}_1$ is equivalent to exchanging the sites at which spins ${\bf S}_{{\bf r}_0}$ and ${\bf S}_{{\bf r}_1}$ are located, as well as those with ${\bf S}_{{\bf r}_2}$ and ${\bf S}_{{\bf r}_3}$. This exchange is achieved by applying ${\hat T}$. The application of the matrix ${\hat S} = {\hat T} \cdot {\hat R}$ leaves the spatial spin structure of the system unchanged. Given the definition of the Hamiltonian matrix in momentum space
\begin{equation}
  \begin{split}
    {\hat H}({\bf k}) =
    &\left(
    \begin{array}{cc}
      -\mu \tau_z \otimes \sigma_0 + \Delta \tau_x \otimes \sigma_0  + J \tau_0 \otimes \left({\bf S}_1 \cdot \vec{\sigma} \right) & -t \cos({\bf k} \cdot {\bf a}_1) \tau_z \otimes \sigma_0 \\
      -t \cos({\bf k} \cdot {\bf a}_1) \tau_z \otimes \sigma_0  & -\mu \tau_z \otimes \sigma_0  + \Delta \tau_x \otimes \sigma_0  + J \tau_0 \otimes \left({\bf S}_2 \cdot \vec{\sigma} \right) \\
      -t \cos({\bf k} \cdot {\bf a}_2) \tau_z \otimes \sigma_0 & -t \cos({\bf k} \cdot ({\bf a}_2 - {\bf a}_1)) \tau_z \otimes \sigma_0  \\
      -t \cos({\bf k} \cdot ({\bf a}_2 - {\bf a}_1)) \tau_z \otimes \sigma_0 & -t \cos({\bf k} \cdot {\bf a}_2) \tau_z \otimes \sigma_0
    \end{array}
    \right. \hspace{0.5cm} ... \\  ... \hspace{0.5cm}
    &\left.
    \begin{array}{cc}
      -t \cos({\bf k} \cdot {\bf a}_2) \tau_z \otimes \sigma_0  & -t \cos({\bf k} \cdot ({\bf a}_2 - {\bf a}_1)) \tau_z \otimes \sigma_0  \\
      -t \cos({\bf k} \cdot ({\bf a}_2 - {\bf a}_1)) \tau_z \otimes \sigma_0 & -t \cos({\bf k} \cdot {\bf a}_2) \tau_z \otimes \sigma_0 \\
    -\mu \tau_z \otimes \sigma_0 + \Delta \tau_x \otimes \sigma_0 + J \tau_0 \otimes \left({\bf S}_3 \cdot \vec{\sigma} \right) & -t \cos({\bf k} \cdot {\bf a}_1) \tau_z \otimes \sigma_0 \\
      -t \cos({\bf k} \cdot {\bf a}_1) \tau_z \otimes \sigma_0 & -\mu \tau_z \otimes \sigma_0 + \Delta \tau_x \otimes \sigma_0 + J \tau_0 \otimes \left({\bf S}_4 \cdot \vec{\sigma}\right)
    \end{array}
    \right)
  \end{split}
\end{equation}
it then follows, that
\begin{align}
 {\hat H}({\bf k}) = {\hat S}^\dagger  {\hat H}({\bf k})  {\hat S}
\end{align}
We note that the eigenvalues of ${\hat S}$ are $\pm i$.

In order to block-diagonalize ${\hat H}({\bf k})$, we need to find the unitary transformation that diagonalizes ${\hat S}$. This can be achieved by first finding the unitary transformations ${\hat U}_{{\hat T},{\hat R}}$ that diagonalizing ${\hat T}$ and ${\hat R}$ separately. These are given by
\begin{subequations}
\begin{align}
  {\hat U}_{{\hat T}}&= \frac{1}{\sqrt{2}}
  \begin{pmatrix}
    1 & -1 & 0 & 0 \\
    1 & 1 & 0 & 0 \\
    0 & 0 & 1 & -1 \\
    0 & 0 & 1 & 1 \\
  \end{pmatrix}
  \otimes {\bf 1}_{4\times4} \; ; \\
  {\hat U}_{{\hat R}}&=
  {\bf 1}_{4\times4} \otimes
  \begin{pmatrix}
    \sqrt{\frac{1}{6} (3 + \sqrt{3})} & - \frac{i}{\sqrt{3+\sqrt{3}}} & 0 & 0 \\
    - \frac{i}{\sqrt{3+\sqrt{3}}} & \sqrt{\frac{1}{6} (3 + \sqrt{3})} & 0 & 0 \\
    0 & 0 & \sqrt{\frac{1}{6} (3 + \sqrt{3})} & - \frac{i}{\sqrt{3+\sqrt{3}}} \\
    0 & 0 & - \frac{i}{\sqrt{3+\sqrt{3}}} & \sqrt{\frac{1}{6} (3 + \sqrt{3})}\\
  \end{pmatrix} \; .
\end{align}
\end{subequations}
The unitary transformation ${\hat U}_{\hat S} = \left( {\hat U}_{{\hat T}} \cdot {\hat U}_{{\hat R}} \right) \cdot {\hat P}$, with
\begin{equation}
  {\hat P} =
\left(
\begin{array}{cccccccccccccccc}
 1 & 0 & 0 & 0 & 0 & 0 & 0 & 0 & 0 & 0 & 0 & 0 & 0 & 0 & 0 & 0 \\
 0 & 0 & 0 & 0 & 0 & 0 & 0 & 0 & 1 & 0 & 0 & 0 & 0 & 0 & 0 & 0 \\
 0 & 1 & 0 & 0 & 0 & 0 & 0 & 0 & 0 & 0 & 0 & 0 & 0 & 0 & 0 & 0 \\
 0 & 0 & 0 & 0 & 0 & 0 & 0 & 0 & 0 & 1 & 0 & 0 & 0 & 0 & 0 & 0 \\
 0 & 0 & 0 & 0 & 0 & 0 & 0 & 0 & 0 & 0 & 1 & 0 & 0 & 0 & 0 & 0 \\
 0 & 0 & 1 & 0 & 0 & 0 & 0 & 0 & 0 & 0 & 0 & 0 & 0 & 0 & 0 & 0 \\
 0 & 0 & 0 & 0 & 0 & 0 & 0 & 0 & 0 & 0 & 0 & 1 & 0 & 0 & 0 & 0 \\
 0 & 0 & 0 & 1 & 0 & 0 & 0 & 0 & 0 & 0 & 0 & 0 & 0 & 0 & 0 & 0 \\
 0 & 0 & 0 & 0 & 1 & 0 & 0 & 0 & 0 & 0 & 0 & 0 & 0 & 0 & 0 & 0 \\
 0 & 0 & 0 & 0 & 0 & 0 & 0 & 0 & 0 & 0 & 0 & 0 & 1 & 0 & 0 & 0 \\
 0 & 0 & 0 & 0 & 0 & 1 & 0 & 0 & 0 & 0 & 0 & 0 & 0 & 0 & 0 & 0 \\
 0 & 0 & 0 & 0 & 0 & 0 & 0 & 0 & 0 & 0 & 0 & 0 & 0 & 1 & 0 & 0 \\
 0 & 0 & 0 & 0 & 0 & 0 & 0 & 0 & 0 & 0 & 0 & 0 & 0 & 0 & 1 & 0 \\
 0 & 0 & 0 & 0 & 0 & 0 & 1 & 0 & 0 & 0 & 0 & 0 & 0 & 0 & 0 & 0 \\
 0 & 0 & 0 & 0 & 0 & 0 & 0 & 0 & 0 & 0 & 0 & 0 & 0 & 0 & 0 & 1 \\
 0 & 0 & 0 & 0 & 0 & 0 & 0 & 1 & 0 & 0 & 0 & 0 & 0 & 0 & 0 & 0 \\
\end{array}
\right)
\end{equation}
then does not only diagonalize ${\hat S}$, but also sorts its eigenvalues $\pm i$ into block-diagonal form such that the first 8 states belong to the
eigenvalue $i$ and the last 8 states to $-i$. The Hamiltonian matrix ${\hat H}({\bf k})$ is now block-diagonalized (because the two eigenspaces of
$\hat S$ cannot be mixed) through the transformation
\begin{align}
{\hat U}^\dagger_{\hat S} {\hat H}({\bf k}) {\hat U}_{\hat S} = \left( \begin{array}{cc} {\hat H}^\prime({\bf k}) & 0 \\ 0 & {\hat H}^\prime({\bf k}) \end{array} \right) \; ,
\end{align}
where
\begin{equation}
  \begin{split}
    {\hat H}^\prime({\bf k}) = &\left(
    \begin{array}{cccc}
     \frac{J}{\sqrt{3}}-2 t \cos (k_x)-\mu  & \Delta  & i \sqrt{\frac{2}{3}} J & 0 \\
     \Delta  & \frac{J}{\sqrt{3}}+2 t \cos (k_x)+\mu  & 0 & i \sqrt{\frac{2}{3}} J \\
     -i \sqrt{\frac{2}{3}} J & 0 & -\frac{J}{\sqrt{3}}+2 t \cos (k_x)-\mu  & \Delta  \\
     0 & -i \sqrt{\frac{2}{3}} J & \Delta  & -\frac{J}{\sqrt{3}}-2 t \cos (k_x)+\mu  \\
     -4 t \cos \left(\frac{k_x}{2}\right) \cos \left(\frac{\sqrt{3} k_y}{2}\right) & 0 & 0 & 0 \\
     0 & 4 t \cos \left(\frac{k_x}{2}\right) \cos \left(\frac{\sqrt{3} k_y}{2}\right) & 0 & 0 \\
     0 & 0 & 4 t \sin \left(\frac{k_x}{2}\right) \sin \left(\frac{\sqrt{3} k_y}{2}\right) & 0 \\
     0 & 0 & 0 & -4 t \sin \left(\frac{k_x}{2}\right) \sin \left(\frac{\sqrt{3} k_y}{2}\right)
     \end{array}
     \right. \hspace{0.5cm} ... \\
     &
     ... \hspace{0.5cm} \left.
    \begin{array}{cccc}
     -4 t \cos \left(\frac{k_x}{2}\right) \cos \left(\frac{\sqrt{3} k_y}{2}\right) & 0 & 0 & 0 \\
     0 & 4 t \cos \left(\frac{k_x}{2}\right) \cos \left(\frac{\sqrt{3} k_y}{2}\right) & 0 & 0 \\
     0 & 0 & 4 t \sin \left(\frac{k_x}{2}\right) \sin \left(\frac{\sqrt{3} k_y}{2}\right) & 0 \\
     0 & 0 & 0 & -4 t \sin \left(\frac{k_x}{2}\right) \sin \left(\frac{\sqrt{3} k_y}{2}\right) \\
     -\frac{J}{\sqrt{3}}-2 t \cos (k_x)-\mu  & \Delta  & \sqrt{\frac{2}{3}} J & 0 \\
     \Delta  & -\frac{J}{\sqrt{3}}+2 t \cos (k_x)+\mu  & 0 & \sqrt{\frac{2}{3}} J \\
     \sqrt{\frac{2}{3}} J & 0 & \frac{J}{\sqrt{3}}+2 t \cos (k_x)-\mu  & \Delta  \\
     0 & \sqrt{\frac{2}{3}} J & \Delta  & \frac{J}{\sqrt{3}}-2 t \cos (k_x)+\mu
    \end{array}
    \right)
  \end{split}
\end{equation}

\newpage

\begin{center}
{\bf Section 3: Topological phase diagram for the experimentally motivated model}
\end{center}

\begin{figure}[h]
  \centering
  \includegraphics[width=12.5cm]{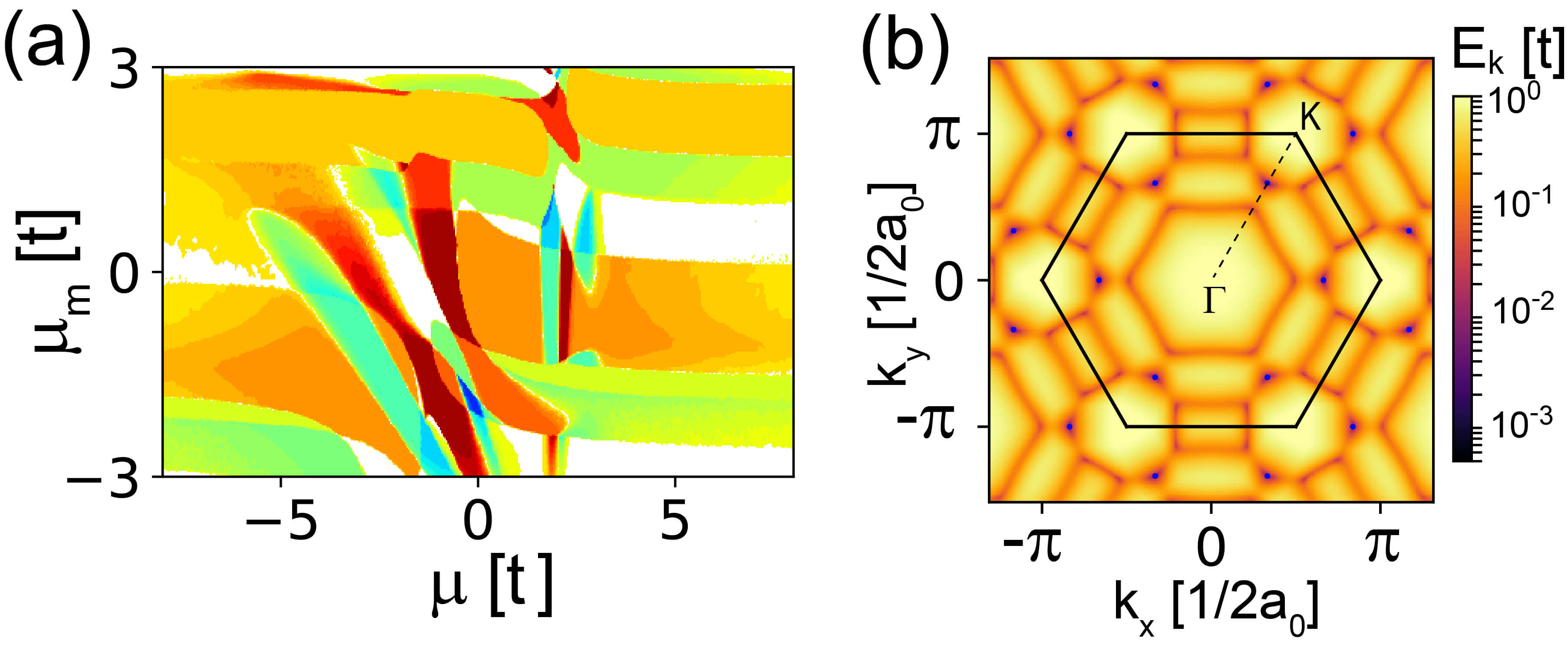}
  \caption{(a) Topological phase diagram in the $(\mu, \mu_m)$ plane with $JS=t$ and $t_{hyb}=0.5t$. Parameters are $\Delta=0.3t$ and $t_m=t$. (b) Plot of the dispersion $E_{\bf k}$ of the lowest energy band at the phase transition between the $C=12$ and $C=0$ phases.}
  \label{fig:SI_Fig2}
\end{figure}

Fig.~4(b) of the main text shows that topological phase diagram for the experimentally motivated model of Eq.(5) in the main text in the $(\mu, JS)$-plane. To demonstrate the ubiquity of topological phases in parameter space, we present in Fig.~\ref{fig:SI_Fig2}(a) the topological phase diagram in the $(\mu,\mu_m)$-plane.
This result confirms that topological phases exist over a wide range of parameters, and are thus a robust phenomenon associated with the 3Q magnetic structure. Finally, in Fig.~\ref{fig:SI_Fig2}(b) we present a plot of the lowest energy band, $E_{\bf k}$, in the Brillouin zone at the transition between the $C=12$ and $C=0$ phases. In contrast to the results shown in Fig.~1(c) of the main text, the gap closing occurs at non-time-reversal invariant points along the $\Gamma -K(K^\prime)$ lines [see dashed line in Fig.~\ref{fig:SI_Fig2}(b)], whose multiplicity of $m=12$ leads to a change in the Chern number by $\Delta C=12$.